\documentclass[showpacs,twocolumn]{revtex4}

\usepackage{graphicx}
\usepackage{dcolumn}
\usepackage{amsmath}
\usepackage{hyperref}
\usepackage{verbatim}
\begin{document}

\title{On the global visibility of a singularity in spherically symmetric gravitational collapse}
\author{Sanjay Jhingan $^{a}$}
\email{sanjay.jhingan@gmail.com}
\author{Sahal Kaushik $^{b}$}
\email{sahal.kaushik@gmail.com}

\affiliation{$^{a}$ Centre for Theoretical Physics, Jamia Millia Islamia,
New Delhi 110025, India}
\affiliation{$^{b}$Department of Physics, Indian Institute of Technology, Kanpur 208016, India}


 \begin{abstract}
We revisit the gravitational collapse of spherically symmetric Lema\^itre - Tolman - Bondi (LTB) dust models. A sufficient condition for global visibility of singularity is given. This condition also allows us to extend the condition of local visibility to mass functions which are not Taylor expandable near the centre. 
 \end{abstract}


\maketitle

\section{Introduction}

The gravitational collapse of a spherically symmetric ball of dust by Oppenheimer and Snyder (OS) gives the most transparent demonstration of formation of blackholes in general relativity \cite{OPS}. The key idea of an event-horizon hiding the singularity at the center of collapsing star is elucidated by infinite redshift of the signals from its boundary as seen by an asymptotic observer, far away from the boundary. Thus, an event-horizon provides a protection from a singularity. However, a singularity does raise an important question about viability of the general theory of relativity, and covering it by a horizon only reduces it to a minor embarrassment for the theory. Earlier, singularities were seen as artefact of underlying symmetry of the collapsing scenario, as the collapsing dust particles were aimed at the center. It was believed that any departure from spherical symmetry would make singularities disappear. However, two major developments in the 60's changed our perception on singularities. First, the observation of cosmic microwave background which strongly supported the idea that universe had a singular beginning, and second a series of theorems by Hawking and Penrose indicating that singularities can occur under very generic conditions both in cosmology and gravitational collapse of stars \cite{hw}. This essentially meant that singularities could not be swept under a carpet (event horizon) but, in a way, should be stitched to it.

In the late $60s$, the OS model led to blackhole formation as the "establishment view-point" in classical general relativity, if the remnant mass of the collapsing star was more than a few solar masses \cite{HWIS}. However, with the proof of singularity theorems there was an increasing need to protect general relativity, an otherwise very successful classical theory, from such a catastrophic consequence. Most of these attempts led to various conjectures shielding the singularity. Two of the most widely discussed proposals were the cosmic censorship conjecture by Penrose \cite{pen-ccc}, and the Hoop conjecture by Thorne \cite{hoop}. Till date both remain unproven, and the Cosmic censorship conjecture is considered as one on the most important unsolved problem in classical general relativity (see \cite{revccc} for reviews). 

With progress in our understanding of solutions to Einstein equations the evidence against cosmic censorship has only mounted. Nevertheless most of these counter-examples have been limited to spacetimes with assumptions on underlying symmetry (spherical/cylindrical) and choice matter models (dust/directed radiation/perfect fluids etc.) \cite{various, christ, newman, dwivedi, joshisingh, HM1, H2, HM2}. With advent of string/M theory as the best candidate theory for quantum gravity it is considered likely that in late stages of collapse we have higher order curvature corrections to Einstein equations. However, several studies initiated in this area have been unable to restore censorship \cite{maeda, sgj, tetsuya}. 

The analysis of singularities in almost all the cases mentioned above has been limited to radial null geodesics in the neighbourhood of singularity. This violates only the so called week censorship conjecture. The case of astrophysical interest would be a violation of strong censorship which requires the singular geodesic to reach boundary of the collapsing cloud without getting trapped.  This can allow for a possibility to model extreme high energy phenomenon (e.g. gamma ray bursts) on naked singularities. In this paper we derive a sufficient condition for existence of a globally visible singularity forming in dust collapse. 
  
In the next Section we give a brief overview of marginally bound dust models. This is followed by a Section on causal structure of singularity arising in gravitational collapse. The main results are summarised in the Section on local and global visibility of singularity. We end with a Section on discussion and concluding remarks. 
 
\section{The model}
The well known Lema{\^i}tre-Tolman-Bondi (LTB) spacetime \cite{LTB} is a spherically symmetric solution of Einstein field equations with metric of the form 
\begin{equation}\label{spti}
	ds^2 = -dt^2 + \frac{(R')^2}{1+f(r)} dr^2 + R^2 (d\theta^2 + \sin^2\theta d \phi^2) .
\end{equation}
It is sourced by an energy-momentum tensor in the form of a pressure free perfect fluid (equation of state $p=0$), given by
\begin{equation}\label{emt}
	T^{\mu\nu}= \epsilon \delta^{\mu}_t \delta^{\nu}_t . 
\end{equation} 
Here, energy density $\epsilon$,  and area coordinate $R$ are functions of $r$ and $t$. As a reasonability condition on the energy-momentum tensor in GR we impose some energy condition on initial data. We impose weak energy condition which in case of dust  restricts energy density to positive values. The underlying symmetry of spacetime allows us to define an invariant mass function $2 m = R (1-g^{\mu \nu} \partial_{\mu}R \, \partial_{\nu} R)$, giving mass inside a sphere of radius $R$. The Einstein field equations take a particularly simple form in terms of the mass function:
\begin{eqnarray} \label{eq1}
	\epsilon(r,t) & = & \frac{m'}{4\pi R^2 R'},\\ \label{eq3}
	{\dot R}^2 &=& \frac{2 m}{R} + f(r),
\end{eqnarray}
where a superscript "prime" and an "over-dot" signify partial derivatives with respect to $r$ and $t$, respectively. The first equation is the ${t,t}$ component of the Einstein field equations, and second is the definition of mass function giving dynamics of the collapsing shells. Since dust particles follow geodesics the mass inside a shell is conserved, hence mass function is a function of $r$ only. 

Note that the form of Eq. (\ref{emt}) allows us to interpret $f(r)$ as the total energy function. From the Eq. (\ref{eq3}) describing growth of energy density we see that there are both \emph{shell-crossing} $(R'=0)$, and \emph{shell-focusing} $(R=0)$ singularities in these models. The shell crossing singularities can be easily avoided in these models by choosing appropriate initial conditions, i.e., considering models with decreasing density away from the center, and by not giving any additional inward velocity to the outer shells. The first condition is physically realistic as we expect stars to have maximum density at the center. The marginally bound scenario ensures the second condition as all shells start at rest from infinity. Thus, we will be concerned with shell focusing singularities here. 

The LTB solution for the marginally bound case can be written as 
\begin{equation}\label{soln}
R^\frac{3}{2}= r^\frac{3}{2} - \frac{3}{2} \sqrt{2m}\; t \;.
\end{equation}
This is the integrated form of Eq. \ref{eq3}, and where we have fixed the integration constant using scaling freedom $R(0,r) = r$, i.e., on initial hypersurface we equate the proper and the coordinate distance. In what follows we shall analyse the marginally bound models ($f(r)=0$) only.

\section{Singularity and Horizons}

As discussed in previous section, in dust spacetime, the shell crossing singularities can be easily avoided by a suitable choice of initial data. From Eq. (\ref{soln}), the singularity curve describing a shell focusing singularity $R=0$, in spacetime is
\begin{equation}\label{singcurve}
	t_0(r) = \frac{2r^{3/2}}{3\sqrt{2m}} .
\end{equation}
Since a physically reasonable density profile is a monotonic decreasing function away from the center, the singularity forms at the center first. Also, this restricts the spacetime coordinates to values
\[
0\leq r \leq r_c , \quad -\infty < t < t_0(r) ,
\]
where $r_c$ is the boundary of the collapsing cloud.  

The characteristic feature of a black hole is that it represents a region of spacetime from where no information (matter or energy) can escape. This phenomenon is a consequence of the fact that propagation of light is influenced by gravitational field. If the gravitational field of an object is strong enough  the light cannot escape eventually falling back on it. This allows us to define a wavefront which is moving outwards yet decreasing in its area, a criterion for light being trapped. More precisely can we define a trapped surface as "a closed (compact without boundary) spacelike 2-surface which has the property that the null geodesics which meet this surface orthogonally are all converging in its neighbourhood" \cite{penrose-trap}. 

The equation for outgoing radial null geodesics in the marginally bound dust spacetime is   
\begin{equation}\label{rngs}
\frac{dt}{dr} = \frac{\partial R}{\partial r}.
\end{equation}
Using outgoing null geodesic equation we can derive the following expression for a change in the proper distance $R$ as
\begin{equation}
\frac{dR}{dr} = \left(1-\sqrt{\frac{2m}{R}}\right) R' .
\end{equation}
Since we are assuming no shell-crossings ($R' > 0$) everywhere in the spacetime, $dR/dr$ is either positive, negative or $0$, when $2 m > R$, $2m < R$ or $2m =R$, respectively.  Therefore, the surface defined by
\begin{equation}\label{apparent}
R = 2m,
\end{equation}
describes a locus of the turning points of out-going radial null geodesics. This is the equation for an apparent horizon. Thus, a second condition for a physically reasonable initial data is absence of trapped surfaces ($2m < r$) everywhere in the collapsing cloud. 

It has been shown by several authors \cite{various} that dust collapse models lead to violation of cosmic censorship conjecture. This revived a wide interest in the problem and in past two decades we have seen censorship violation in various collapse scenarios. Essentially, most of the analytical approaches in this area analyse null geodesics in the neighbourhood of the singularity. What is shown is that depending on initial data we can have radial null geodesics in the spacetime with their past endpoints at singularity. Due to the vacuum exterior the event and the apparent horizons meet at the boundary. When traced backwards the apparent horizon meets singularity at the center. However, tracing the event horizon requires integrating the radial null geodesic equation. It is only in the homogeneous dust collapse model, studied by Oppenheimer and Snyder \cite{OPS}, that one can draw the event horizon curve. 

In what follows we give a brief review of results on local visibility of singularities in dust collapse \cite{dwivedi, psjbook, Barve}. Consider outgoing radial null geodesics in the spacetime (\ref{rngs}). The geodesics with past end points at singularity $t_0(0)$, take the following approximate form:
\begin{equation}\label{approxgeo}
	t = t_0(0) + X r^{\alpha} \;,
\end{equation}
near center. Here $\alpha$ and $X$ are both positive for geodesics to exist in the spacetime. Using solution (\ref{soln}) we expand $R'$ near the center, and evaluate $dt/dr$ using eqn. (\ref{approxgeo}) above. Thus we can write null geodesic equation (\ref{rngs}) near the center where left hand side has unknown parameters $X$ and $\alpha$ and its right hand side has initial data in terms of mass (or energy density) expanded near $r \sim 0$. Thus, existence or otherwise of a naked singularity, which requires positive definite values for $X$ and $\alpha$ gets determined in terms of initial data. In marginally bound dust collapse if the leading inhomogeneity term in the density near the center is either linear or quadratic the singularity is always naked (locally) \cite{Barve}. If it is cubic we have both black hole and naked singularity \cite{Barve, jhingan-cqg}

\section{local and global visibility}\label{sec:global_visibility}

The role of apparent horizon in describing final fate of gravitational collapse and its relation to initial data in dust spacetimes was shown in \cite{jhingan-cqg}. The analytical work in this area has been limited to only local analysis of geodesics surrounding central singularity. An astrophysically interesting scenario would be to have a globally visible singularity \cite{haradajhingan}.  Combining Eqs. (\ref{soln}), (\ref{singcurve}) and (\ref{apparent}) we can write
\begin{equation}\label{singapp}
t_0(r) - t_{ah} (r) =  \frac{4}{3} m(r) .
\end{equation}
Since regular initial data ensures mass should vanish at the center, singularity curve and apparent horizon curve meet there. Whereas the positivity of mass for any non-central shell ($r>0$) ensures that all the non-central points on the singularity curve are safely trapped.  Therefore, it is only the central shell focusing singularity from where light may escape getting trapped. In what follows we shall focus on the central singularity ($r=0$). 

In marginally bound dust collapse initial data can be specified in terms on local expansion of density near central singularity (\ref{eq1}), 
\begin{equation}
\rho(r) = \epsilon(r,0) = \rho_0 + \frac{1}{n!}\rho_n r^n + \cdots =  \frac{1}{4\pi r^2}\frac{dm}{dr}
\end{equation} 
and the corresponding mass function takes the form 
\begin{equation}
m(r) = m_0 r^3 + m_{n} r^{n+3}  + \cdots 
\end{equation}
here $n\geq 1$ takes integral values. As far as local visibility is concerned it is the first non-zero value of $n$ which decides the end-state of collapse \cite{christ, newman, dwivedi, joshisingh, Barve} (see \cite{psjbook} for a review). If the first non-zero term near the center is $n=1$ or $2$, the singularity is always visible. At $n=3$ we can have a transition from visible singularity to a blackhole depending on initial data (central density and inhomogeneity). For $n\geq 4$ the collapse always ends in a blackhole.  However, the analysis provides a sufficient condition for the local visibility of singularity as it is limited to conditions near the center. In the remaining paper we have developed a sufficient condition for the global visibility of singularity.  

In what follows it is convenient to define a auxiliary variable so that $r = r(f)$. The equation of outgoing null geodesics (\ref{rngs}) can be rewritten as 
\begin{equation}\label{nrng}
\frac{dt}{df} = \frac{\partial R}{\partial f}
\end{equation}
Due to absence of any pressure the dust particles follow geodesics, and this conserves mass within a co-moving radius. Thus, instead of comoving distance $r$, we can also use mass function $m(r)$ as a coordinate.   Let $f(r) = (9m(r)/2)^\frac{1}{3}$, be the definition of the auxiliary variable. In terms of $f$ the dust solution (\ref{soln}) can be re-written as 
\begin{equation}
R^\frac{3}{2} = r^\frac{3}{2} - t f^\frac{3}{2} .
\end{equation}
This solution takes a particularly simple form in terms of time of formation of singularity $t_0 = (r/f)^\frac{3}{2}$,
\begin{equation}
R = f (t_0 - t)^\frac{2}{3}  .
\end{equation}
Further, defining $(t_0 - t) = u^3$, we can re-write the radial null geodesic equation Eq. (\ref{nrng}) as
\begin{equation}
\frac{du}{df} = \frac{q}{3u^2} - \frac{1}{3} - \frac{2 q f}{9u^3} .
\end{equation}
where $(dt_0/df) = q$. We make one final transformation:
\begin{equation}\label{defv}
v = \frac{u}{f}  .
\end{equation}
The radial null geodesics in $(v,f)$ coordinates take the following form 
\begin{equation}\label{rngvf}
\frac{dv}{df} = \frac{1}{f}\left[\left(-v-\frac{1}{3}\right) + \frac{q}{f^2}\left(\frac{1}{3v^2} - \frac{2}{9v^3}\right)\right]
\end{equation}

The proper distance $R$, and the comoving coordinates ($t,r$),  have following functional form in the new coordinates $(v,f)$,  
\begin{eqnarray}
R &=& fu^2 = f^3v^2 ,\nonumber \\
r &=& R(t = 0) = f t_0^\frac{2}{3} , \\ \nonumber
t &=& t_0 - u^3 = t_0 - f^3v^3 .
\end{eqnarray}
Since $f\propto m^{1/3}$, for small $f$, $r$ is proportional to $f$ and increases with it. $t$ decreases with $v$, but lines of constant $v$ are not lines of constant $t$, 
\begin{equation}
\left.\frac{\partial t}{\partial f}\right|_v = q-3f^2v^3 .
\end{equation}
The singularity is at $t = t_0$, i.e. , 
\begin{equation}
u = f v = 0 .
\end{equation} 
This definition is crucial in distinguishing regular and singular points along null geodesics. Along a radial null geodesic if $v$ attains a finite value in the limit of approach to center it implies the geodesic is singular; as $f$ tends to zero $u$ approaches $0$ ($t\rightarrow t_0$)  and the singularity is approached. On the other hand, $v$ necessarily blows up if the null geodesic approaches a regular center. In a special case, when $v$ is proportional to $f^{-1}$, $u$ approaches a non-zero value, and a regular center is approached and $v$ blows up.

The apparent horizon is at $R = 2m$, i.e., 
\begin{equation}
v = \frac{2}{3} .
\end{equation}
Any region of spacetime where $v>2/3$ is free of trapped surfaces. If a geodesic crosses the apparent horizon it has to fall back into the singularity, i.e. after crossing apparent horizon $v$ remains bounded in interval (0,2/3). Therefore, for a singular null geodesic to escape to infinity, $v$ has to remain greater than $2/3$ on it throughout the collapsing cloud. In what follows we derive a sufficient condition on a radial null geodesic for which $v$ remains above $2/3$ till boundary of the cloud, and it remains finite in the  $f=0$ limit (i.e., it is singular).  Existence of a such a geodesic ensures that singularity is globally visible.  

From Eq. (\ref{rngvf}), if $v>\frac{2}{3}$, 
\begin{equation}
\frac{dv}{df} = \frac{1}{f}\left[\left(-v-\frac{1}{3}\right) + \frac{q}{f^2}\left(\frac{1}{3v^2} - \frac{2}{9v^3}\right)\right] > 0 ,	
\end{equation}
iff
\begin{equation}\label{condition}
 \frac{q}{f^2} > \frac{3v^3 (3v + 1)}{3v-2} .
\end{equation}
The function ${(3v^3 (3v + 1))}/{(3v-2)}$ has a local minimum $(52 + 30\sqrt3)/9$ at $v= (1/3 + 1/\sqrt{3})$. It approaches $\infty$ as $v\to {2}/{3}$, and as $v\to \infty$. Thus, if ${q}/{f^2} > (52 + 30\sqrt3)/9$ for all $f$, ${dv}/{df} > 0$ in an interval of $v$ containing $({1}/{3} + {1}/{\sqrt3})$.

A null geodesic is characterised by parameters $v$ and $f$. Consider a point with finite values $v>2/3$ and $f$ on it. The geodesic is now extended from this point in both directions. If $q(f)$ satisfies $\frac{q}{f^2} > \frac{1}{9}(52 + 30\sqrt3)$ for all $f$, the geodesic cannot cross the line $v = {1}/{3} + {1}/{\sqrt3}$ again as ${dv}/{df} > 0$ on that line. Therefore, $v$ approaches a finite value (${2}/{3}$) as $f\to 0$, and it remains greater than $2/3$ as $f\to\infty$. This geodesic starts from the singularity at the center, and escapes boundary without getting trapped. 

The sufficient condition for the global visibility of singularity (\ref{condition}) can also be written as
\begin{equation}\label{condapphor}
\frac{dt_0}{dm} > \frac{1}{3}(26+15\sqrt3)
\end{equation}
for all $m$. Alternatively, using Eq. (\ref{singapp}) the global visibility condition reads
\begin{equation}
\frac{dt_{ah}}{dm} > \frac{1}{3}(22+15\sqrt3) 
\end{equation}
for all $m$.

\section{special cases}
In this section we provide some explicit examples of globally visible singularity based on the condition derived above (\ref{condapphor}). 

\subsection{Non-Selfsimilar Collapse}
\begin{itemize}
\item \underline{$\rho = \rho_0(1-r)$} : Consider a collapsing dust cloud with central density $\rho_0$, and comoving boundary $r_b = 1$. Thus region interior to the cloud is characterised by a range $0\leq r \leq 1$.  
The mass as a function of radius for this density profile is 
\begin{equation}
m(r) = 4M\left(r^3 - \frac{3}{4}r^4\right)
\end{equation}
Where $M = m(1)$ is the total mass. This mass profile ensure regular mass function at the center. Another regularity condition on requirement of no trapped surfaces on an initial hypersurface ($2m/r<1$) restricts total mass in the range $ 0 < M < 0.4746 $.
From Eq. (\ref{singcurve}), the singularity curve is
\begin{equation}
t_0 = \frac{2}{3\sqrt{2M}}\frac{1}{\sqrt{4-3r}},
\end{equation}
and 
\begin{equation}
\frac{dt_0}{dm} = \frac{1}{12\sqrt 2}\frac{1}{M^\frac{3}{2}}\frac{1}{r^2(1-r)(4 - 3r)^\frac{3}{2}} .
\end{equation}
The function $r^2(1-r)(4-3r)^\frac{3}{2}$ goes to $0$ as $r$ goes to $0$ or $1$, it has a maximum of 0.0618508 at $r = 0.514191$. Thus
\begin{equation}
\frac{dt_0}{dm} \geq \frac{1}{8.3971}M^{-\frac{3}{2}},
\end{equation}
and the sufficient condition for global visibility (\ref{condapphor}) is satisfied if 
\begin{equation}
M < 0.036149 \;.
\end{equation}

\item \underline{$\rho = \rho_0(1-r^2)$} : The mass function in this case becomes,
\begin{equation}
m(r) = \frac{5}{2}M\left(r^3-\frac{3}{5}r^5\right).
\end{equation}
The condition for absence of trapped surfaces on initial hypersurface gives following condition on initial data, $ 0 < M < 0.48$.
The singularity curve for this profile is
\begin{equation}
t_0 = \frac{2}{3\sqrt{M}}\frac{1}{\sqrt{5-3r^2}} ,
\end{equation}
and 
\begin{equation}
\frac{dt_0}{dm} = \frac{4}{15}\frac{1}{M^\frac{3}{2}}\frac{1}{(5-3r^2)^\frac{3}{2}r(1-r^2)} .
\end{equation}
The function $r(1-r^2)(5-3r^2)^\frac{3}{2}$, goes to $0$ as $r$ goes to $0$ or $1$, and has a maximum vakue $0.206295$ at $r=0.403231$. Thus,
\begin{equation}
\frac{dt_0}{dm} \geq \frac{1}{8.64918}M^{-\frac{3}{2}},
\end{equation}
and the condition for global visibility (\ref{condapphor}) is satisfied if 
\begin{equation}\label{globr2}
M < 0.035443 .
\end{equation}
The globally visibility of singularity for various values of initial mass functions is shown in Fig (\ref{rho2geo}) below. We start with some value $v>2/3$ on the boundary on a radial null geodesic Eq. (\ref{rngvf}) and integrate it towards the center for different values of total mass of the cloud. When the value of total mass violates the global visibility condition (\ref{globr2}), $v$ blows up as we move towards the center showing that such geodesics can only originate from a regular center. 

\begin{figure}[ht]
  \begin{center}
    \includegraphics[height=2.5in]{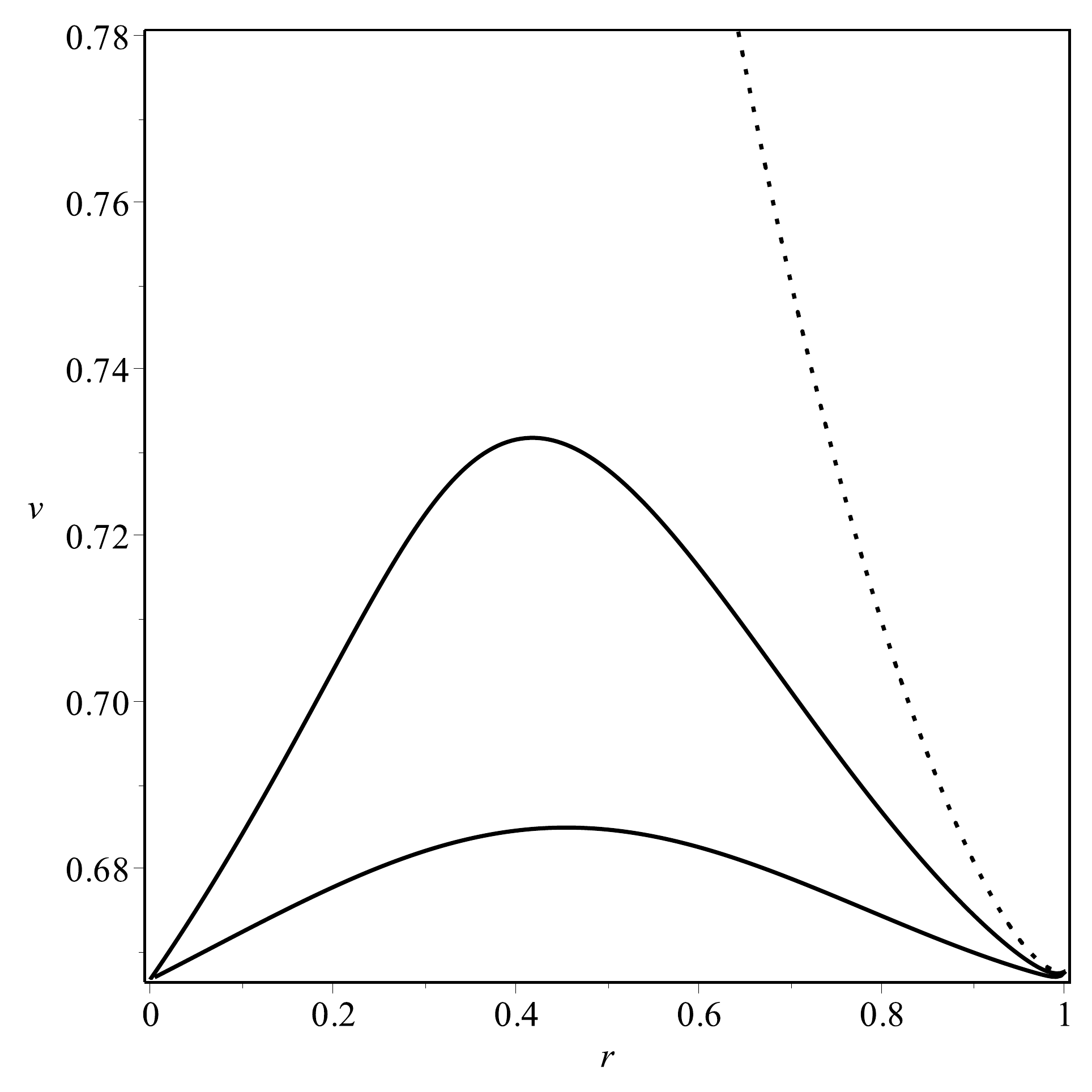}
  \end{center}
  \caption{\small Figure showing regular (dotted) and singular (bold) curves, respectively for $M = 0.05, 0.02$ and $0.01$ }.
  \label{rho2geo}
\end{figure} 

\item \underline{$\rho = \rho_0(1-r^3)$} : The mass function corresponding to this density profile is
\begin{equation}
m = M(2r^3 - r^6) .
\end{equation}
Absence of trapped surfaces on a regular initial hyper surface gives following bound on total mass $0 < M < 0.4835$. Singularity curve and its variation with respect to mass function is given by, respectively,
\begin{equation}
t_0 = \frac{2^{1/2}}{3}\frac{1}{(2 - r^3)^{1/2}} ,
\end{equation}
and 
\begin{equation}
\frac{dt_0}{dm} = \frac{2^{1/2}}{12}\frac{1}{M^\frac{3}{2}}\frac{1}{(1-r^3)(2-r^3)^{\frac{3}{2}}} .
\end{equation}
This has a minimum at $r=0$. This is unlike previous two cases where denominator had a maximum value in the interval $r\in [0,1]$. Thus,
\begin{equation}
\frac{dt_0}{dm} \geq \frac{1}{24}M^{-\frac{3}{2}}
\end{equation}
and the condition (\ref{condapphor}) is satisfied if 
\begin{equation}
M < \frac{1}{4}(26-15\sqrt 3)^\frac{2}{3}  = 0.017949 ,
\end{equation}
and a black hole forms if $M > 0.017949$. 

First, this is same condition as obtained by \cite{Barve}. Assuming $2m = r^3(F_0 + F_3 r^3 + ...)$ the singularity curve is given by: 
\[
t_0 = \frac{2}{3}\frac{r^\frac{3}{2}}{\sqrt{2m}} = \frac{2}{3\sqrt{F_0}} -\frac{2}{3}\frac{F_3}{F_0^\frac{5}{2}}m + \cdots
\]
At $m=0$,
\[ 
\left.\frac{dt_0}{dm}\right|_{m=0} = \frac{2}{3}\frac{-F_3}{F_0^\frac{5}{2}}
\]
So the condition for local visibility in the $n=3$ case is
\[
\left.\frac{dt_0}{dm}\right|_{m=0} > \frac{1}{3}(26+15\sqrt 3)
\]
i.e.,
\[
\frac{-F_3}{F_0^\frac{5}{2}} > \frac{1}{2}(26+15\sqrt 3) = 25.9904.
\] 

For the $\rho (1-r^3)$ profile, $dt_0/dm$ has a minimum at $m=0$. If the singularity is locally visible, $\frac{dt_0}{dm} > \frac{1}{3}(26+15\sqrt 3)$, this condition holds for all $m$, and The singularity is also globally visible.

Therefore, in this case, we have either a black hole or a globally naked singularity, there is no mass such that the singularity is locally but not globally naked. The special nature of this profile comes out in the plot of null geodesics. The singular geodesics in fig. (\ref{rho3geo}) meet the central singularity with different values of $v$ unlike previous two cases where $v=2/3$ at the center on singular geodesics. 
\begin{figure}[ht]
  \begin{center}
    \includegraphics[height=2.5in]{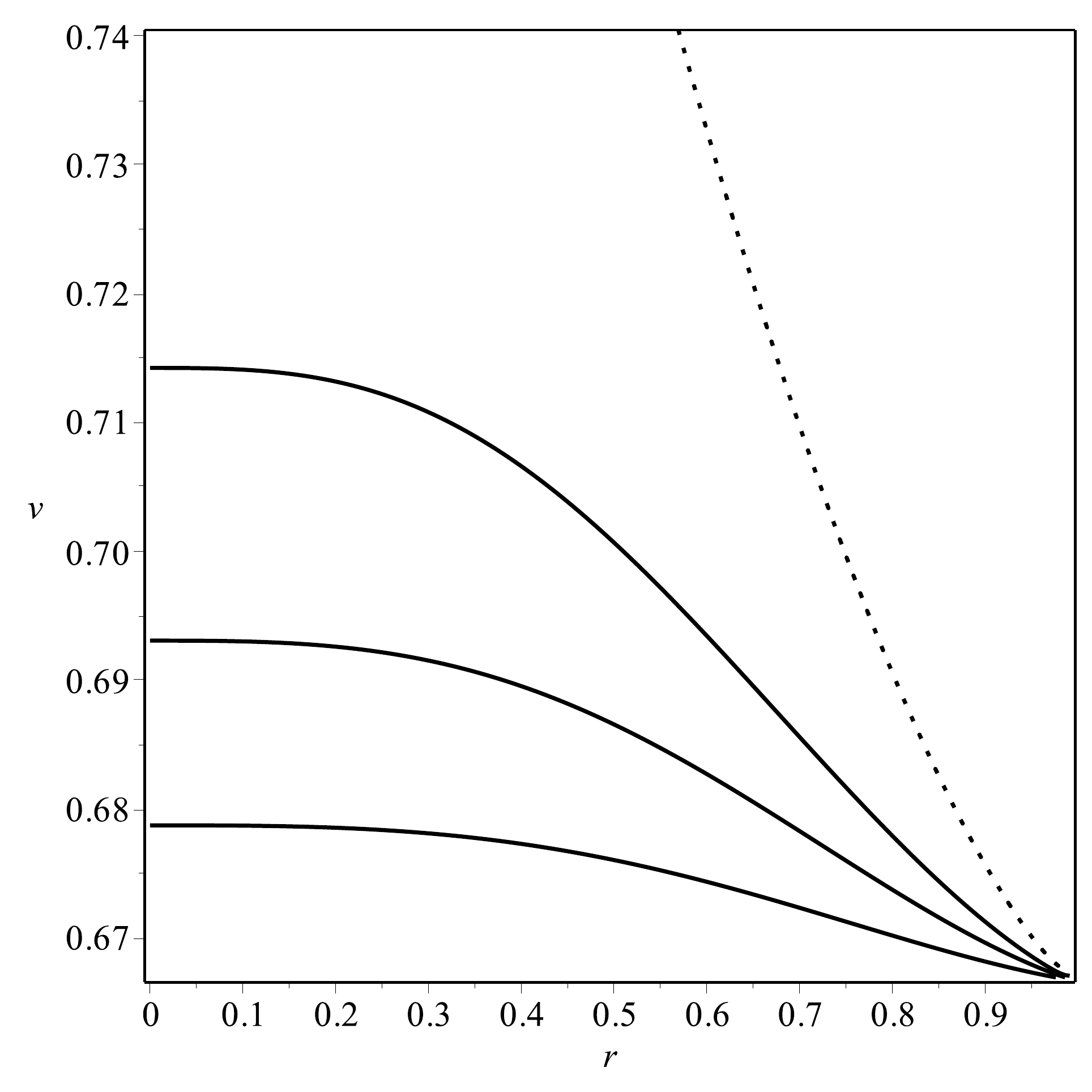}
  \end{center}
  \caption{\small Figure showing a regular (dotted) and singular (bold) curves, respectively, for $M = 0.019, 0.011, 0.008$, and $0.005$ }.
  \label{rho3geo}
\end{figure} 

\end{itemize}

\subsection{Selfsimilar Collapse}
\begin{figure}[ht]
  \begin{center}
    \includegraphics[height=2.5in]{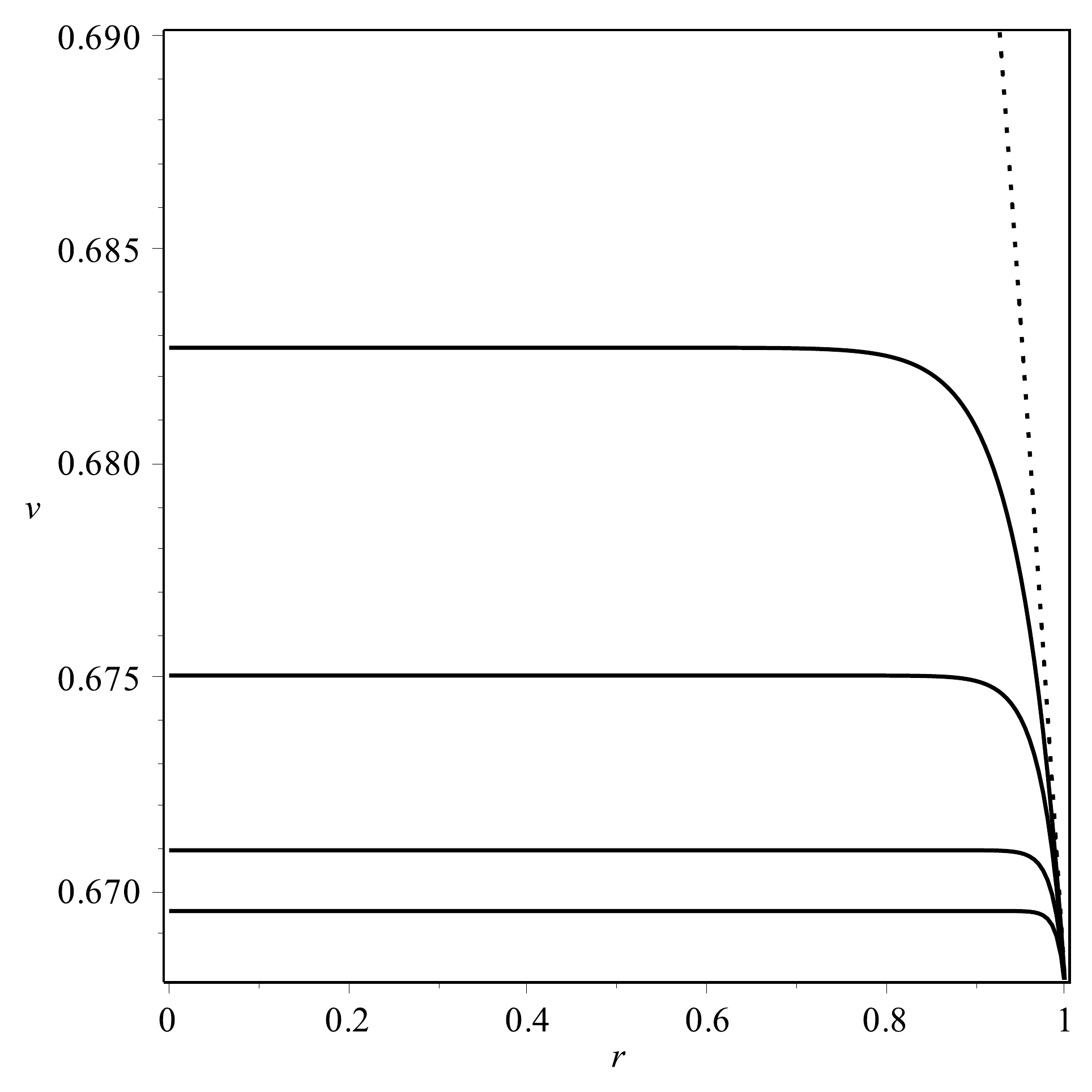}
  \end{center}
  \caption{\small Figure showing a regular (dotted) and singular (bold) curves, respectively, for $\kappa = 0.1, 0.03, 0.02, 0.013 $ and $0.01$ }.
  \label{rhoselfgeo}
\end{figure}

The assumption of self-similarity restricts mass function to have a linear form 
\begin{equation}
m = \kappa r .
\end{equation}
Here $\kappa >0$, and $r=1$ characterizes boundary of the cloud. The condition for absence of trapped surfaces ($2m/r <1$) implies $0 < \kappa < 0.5$. The singularity curve is   
\begin{equation}
t_0 = \frac{2}{3}\frac{r}{\sqrt{2}\kappa} ,
\end{equation}
and its derivative with respect to mass
\begin{equation}
\frac{dt_0}{dm} = \frac{\sqrt 2}{3} \frac{1}{\kappa^\frac{3}{2}} .
\end{equation}
By Eq. (\ref{condapphor}) the singularity is globally naked if
\begin{equation}
\kappa < (26-15\sqrt 3)^\frac{2}{3} 2^\frac{1}{3} = 0.090458
\end{equation}
A black hole forms if $\kappa>0.090458$ ( \cite{barveqm}). In this case also, our sufficient condition is also a necessary condition. Moreover, the nature of singular geodesics, fig. (\ref{rhoselfgeo}),  is very similar to the previous case where they met central singularity with finite but different values of $v$. 

\subsection{A Special Case}
The analysis of the local visibility of the singularity has been limited to density functions which are Taylor expandable near the center. The visibility or otherwise of the singularity is then shown to be related to the first non-zero coefficient in the expansion of density near the center. Clearly, during late stages of collapse, it difficult to justify mathematical properties like smoothness of functions which allows expansion inside the matter cloud.  In what follows we propose as an example, a density profile which is not Taylor expandable near the center. This function has a reasonable behaviour as density falls away from the center, and mass function vanishes at the center. 

Consider an initial density profile of the form
\[
\rho = \rho_0(1+r\ln r - r) .
\]  
In this distribution, the density cannot be expanded as a Taylor series about
the center. The corresponding mass function takes a form
\begin{equation}
m = M(16r^3 - 15r^4 + 12r^4\ln r).
\end{equation}
And, 
\begin{equation}
t_0 = \frac{2}{3\sqrt {2M}}\frac{1}{\sqrt{16 - 15r + 12r\ln r }}
\end{equation}
\begin{equation}
\frac{dt_0}{dm} = \frac{1}{48\sqrt
2}\frac{1}{M^\frac{3}{2}}\frac{1-4\ln r}{r^2(1+r\ln r - r)(16 - 15r +
12r\ln r)^\frac{3}{2}}
\end{equation}
\begin{equation}
\frac{dt_0}{dm} \geq 0.13777 M^{-\frac{3}{2}}
\end{equation}
Condition (\ref{condapphor}) is satisfied if
\begin{equation}
M < 0.039837
\end{equation}
A naked singularity forms if the radius of the cloud is less than
12.55 times the Schwarzschild radius.\\

\begin{figure}[ht]
  \begin{center}
    \includegraphics[height=2.5in]{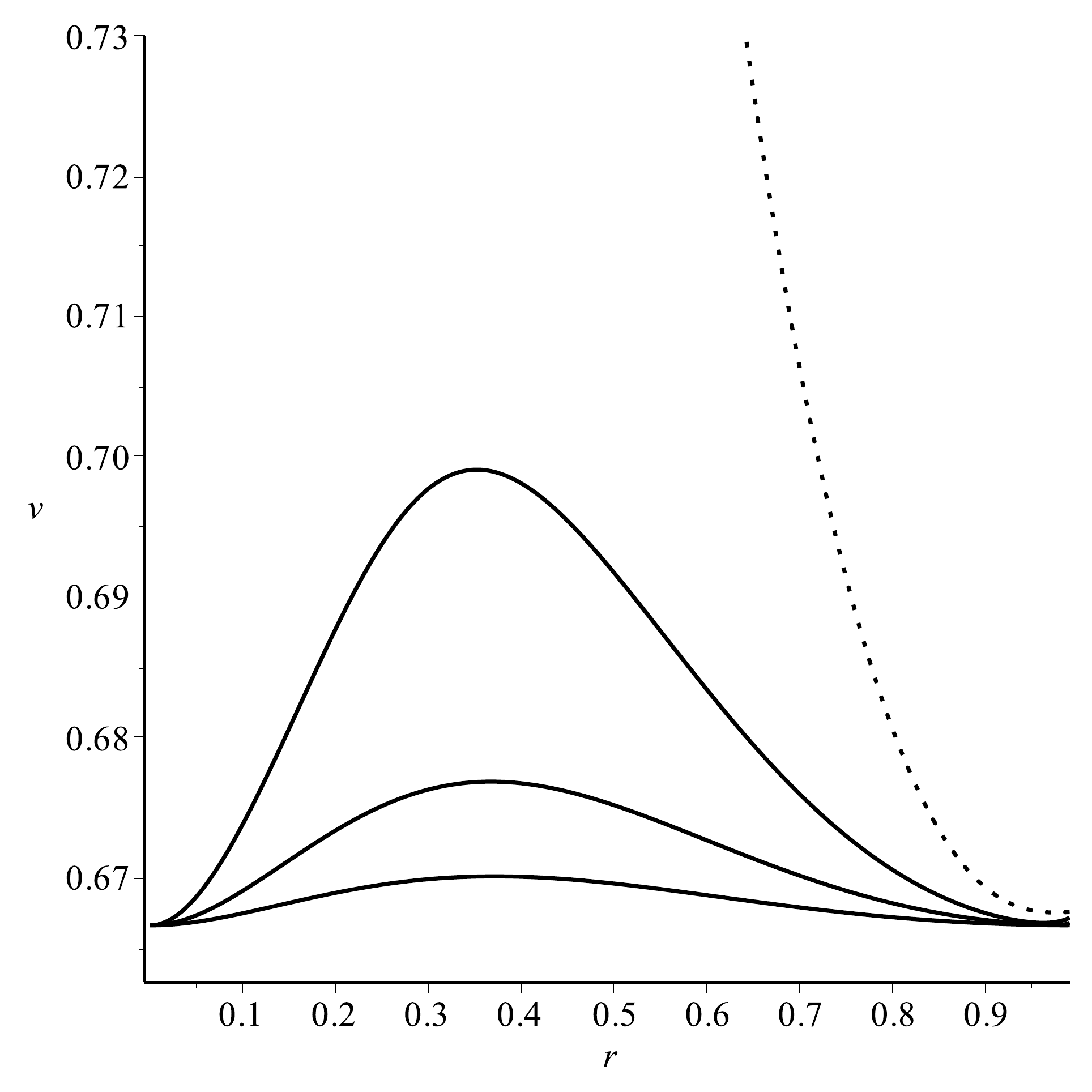}
  \end{center}
  \caption{\small Figure showing a regular (dotted) and singular (bold) curves, respectively, for $M = 0.1, 0.02, 0.01$, and $0.005$ }.
  \label{logrhogeo}
\end{figure} 

\section{Discussion and Conclusions}
The cosmic censorship conjecture remains one of the most important unsolved problem in classical general relativity. The visibility of a singularity to an observer falling on to a blackhole after it crosses the horizon is benign compared to singular geodesics escaping to far away observers. Though such a violation of censorship conjecture should have serious consequences for physics, it also allows us possibility to model violent high energy phenomenon in astrophysics using naked singularities. Recently, Jhingan et al. \cite{jhi-dwi-bar} proposed a possibility to quantify energy coming out from such singularities.

In this paper we have studied collapse of spherically symmetric dust models. These models have provided some of the most serious counter examples to cosmic censorship conjecture. The studies in dust models have been focused on developing local nakedness criterion bases on expansion of initial data near the center. This restricts the analysis to a special class of initial data space, i.e., expandable near the center. The analysis of causal structure has been limited to analysing apparent horizon curve. Moreover, astrophysically interesting scenarios of globally visible singularities have been limited to numerical studies (see, for example, \cite{miyamoto}).

To see the causal structure of a spacetime we need to trace the null geodesics in it. However, this can be done only in homogeneous dust models. In inhomogeneous models it is not possible to integrate null geodesics analytically and our understanding is based on numerical studies \cite{miyamoto}. In this paper we have given a sufficient criterion for existence of a globally naked singularity (\ref{condapphor}). It is shown how the known examples of initial data which leads to formation locally visible singularities can be extended to global visible cases. This condition for global visibility does not depend on expansion of functions near the center. Therefore, it allows us to probe end-states of gravitational collapse for a wider class of functions representing initial data. It would be interesting to see if this condition for global visibility can yield results in case of perfect fluid collapse where local analysis near singularity relating initial data with visibility of singularity is not known.

\begin{acknowledgments}
SK acknowledges support from the Summer Research Fellowship Program of the Indian Academy of Sciences, Bangalore. SJ acknowledges support under UGC minor research project (42-1068/2013(SR)).
\end{acknowledgments}


\begin{thebibliography}{99}
	\bibitem{OPS} J R Oppenheimer and H. Snyder, Phys. Rev {\bf 56} 455 (1939).
	
	\bibitem{hw} R. Penrose, Phys. Rev. Lett. {\bf 14} 57 (1965); S. W. Howking and R. Penrose, Proc. Roy. Soc.  {bf A 314} 529 (1970); R. Geroch and G. T. Horowitz, in \emph{General Relativity: an Einstein Centenary Survey}, eds. S. W Hawking and W. Israel (Cambridge Univ. Press, Cambridge, 1979).

\bibitem{HWIS} S. W. Hawking, in \emph{Black Holes: Les Houches lectures}, eds. C. Dewitt and B. S. Dewitt (B. S. Amsterdam, North Holland, 1972).
		
	\bibitem{pen-ccc} R. Penrose, Rivista Del Nuovo Cimento, Numero Speciale {\bf 1} 252 (1969) 

	\bibitem{hoop} K. Thorne, \emph{Black Holes and Time Warps: Einstein's Outrageous Legacy}, W. W. Norton \& Company, (1995).

\bibitem{revccc} C. J. S.~Clarke, {Classical
Quantum Gravity} {\bf 10}, 1375 (1993);  R. M.~Wald, gr-qc/9710068; S.~Jhingan and G.~Magli, gr-qc/9903103;   T. P.~Singh, {J. Astrophys. Astron.} {\bf 20}, 221 (1999);  P. S.~Joshi, {Pramana} {\bf 55}, 529 (2000);  T.~Harada, Pramana {\bf 63}, 741 (2004); P. S. Joshi, {\it Gravitational Collapse and spacetime singularities} (Cambridge University Press, Cambridge, England, 2007).
	
	\bibitem{various} D. M. Eardley and L. Smarr, {Phys. Rev. D}
{\bf 43}, 2239 (1979); D. M. Eardley, {\it Gravitation in
Astrophysics, ASI Series}, edited by B. Carter and J. B. Hartle
(NATO Advanced Study Institute, Series B: Physics, Vol )(Plenum
Press, New York, 1986) pp 229-235; B. Waugh and K. Lake, {Phys. Rev.} {\bf D 38}, 1315 (1988); J. P. S. Lemos, {Phys. Lett. A} {\bf 158}, 271 (1991); {\it ibid.} { Phys. Rev. Lett.} {\bf 68}, 1447 (1992);  I. H. Dwivedi and P. S. Joshi, {Class. Quantum Grav.} {\bf 9}, L69 (1992).; P. S. Joshi and T. P. Singh, {Gen. Relativ. Gravit.} {\bf 27}, 921 (1995); {Phys. Rev.} {\bf D 51}, 6778 (1995).; P. S. Joshi and I. H. Dwivedi, {Phys. Rev.}
 {\bf D 47}, 5357 (1993); 
	
	\bibitem{christ} D. Christodoulou, Comm. Math. Phys. {\bf 93} 171 (1984).
     
     \bibitem{newman} R. P. A. C. Newman, Class. Quant. Grav. {\bf 3} 527 (1986).

    \bibitem{dwivedi}P. S. Joshi and I. H. Dwivedi, Phys. Rev. D {\bf 47} 5357 (1993).
    
  \bibitem{joshisingh} P. S. Joshi and T. P. Singh, Class. Quantum Grav. {\bf 13} 559 (1996).
  
 \bibitem{HM1} T. Harada and H. Maeda, Phys. Rev. D{\bf 63} 084022 (2001).
 
 \bibitem{H2} T. Harada, {Class. Quantum Grav.} {\bf 18}, 4549 (2001).
  
  \bibitem{HM2} T. Harada and H. Maeda, {Class. Quantum Grav.} {\bf 21}, 371 (2004).
    
	\bibitem{maeda} H. Maeda, Phys. Rev. D{\bf 73}, 104004 (2006); H. Maeda, M. Nozawa, Phys. Rev. D{\bf 77} 064031 (2008).
	
	\bibitem{sgj} S. Jhingan and S. G. Ghosh, Phys. Rev. D{\bf 81} 024010 (2010); S. G. Ghosh and S. Jhingan, Phys. Rev. D{\bf 82} 024017 (2010); N. Dadhich, S. G. Ghosh and S. Jhingan, Phys. Rev. D{\bf 88} 084024 (2013).
	
	\bibitem{tetsuya} S. Ohashi, T. Shiromizu and S. Jhingan, Phys. Rev. D{\bf 84} 024021 (2011); S. Ohashi, T. Shiromizu and S. Jhingan, Phys. Rev. D{\bf 86} 044008 (2011).
	
	\bibitem{LTB} G. Lema\^itre, Ann. Soc. Sci. Bruxelles I A {\bf 53}, 51 (1933); R. C. Tolman, Proc. Nat. Acad. Sci. {\bf 20} 169 (1934); H. Bondi, Mon. Not. Roy. Astron. Soc. {\bf 107}, 410 (1947).
		
	\bibitem{penrose-trap} R. Penrose, in \emph{Battelle Rencontres} (ed. C. M. De Witt and J. A. Wheeler) (New York 1968); Gen. Rel. Grav. {\bf 34} 1140 (2002) [reprinted].
		
		
	\bibitem{psjbook} P. S. Joshi, \emph{Global Aspects in Gravitation and Cosmology}, Clarendon Press, Oxford (1993). 
	
    \bibitem{Barve} S. Barve, T. P. Singh and C. Vaz and L. Witten,  Class. Quant. Grav. {\bf 16} 1727 (1999).
	
    \bibitem{jhingan-cqg} S. Jhingan, P. S. Joshi and T. P. Singh, Class. Quantum Grav. {\bf 13} 3057 (1996).
    
    \bibitem{haradajhingan} U. Miyamoto, s. Jhingan and T. Harada, Prog. Theor. Exp. Phys. 053E01 (2013) 
		     
      
     \bibitem{barveqm} S. Barve, T. P. Singh and C. Vaz and L. Witten, Phys. Rev. D {\bf 58} 104018 (1998).
     
     \bibitem{jhi-dwi-bar} S. Jhingan, I. H. Dwivedi and S. Barve, Phys. Rev. D{\bf 84} 024001 (2011). 
    
    \bibitem{miyamoto} U. Miyamoto, S. Jhingan and T. Harada, Prog. Theor. Exp. Phys. 053E01 (2013).
     	
\end{thebibliography}
\end{document}